\documentclass[prl,twocolumn,showpacs,amsfonts,amsmath,floatfix]{revtex4}

\usepackage{graphicx}
\usepackage{epstopdf}
\usepackage{amssymb}
\usepackage{amsmath}
\usepackage{xspace}
\usepackage{bbold}

\newcommand{\be}{\begin{equation}}
\newcommand{\ee}{\end{equation}}

\begin{document}

\title {Filtering Random Matrices: The Effect of Incomplete Channel Control in Multiple Scattering}

\author{A. Goetschy}
\email{arthur.goetschy@yale.edu}
\affiliation{Department of Applied Physics, Yale University, New Haven, Connecticut 06520, USA}
\author{A. D. Stone}
\affiliation{Department of Applied Physics, Yale University, New Haven, Connecticut 06520, USA}

\begin{abstract}
We present an analytic random matrix theory for the effect of incomplete channel control on the measured statistical properties of the scattering matrix of a disordered multiple-scattering medium. When the fraction of the controlled input channels, $m_1$, and output channels, $m_2$, is decreased from unity, the density of the transmission eigenvalues is shown to evolve from the bimodal distribution describing coherent diffusion, to the distribution characteristic of uncorrelated Gaussian random matrices, with a rapid loss of access to the open eigenchannels. The loss of correlation is also reflected in an increase in the information capacity per channel of the medium. Our results have strong implications for optical and microwave experiments on diffusive scattering media.
\end{abstract}

\pacs{42.25.Bs, 05.60.Cd, 02.10.Yn}
\maketitle

Describing wave propagation in strongly scattering media is a fundamental challenge in disordered systems theory, relevant to problems in electromagnetism, acoustics and electron transport.  The scattering matrix, which encodes fully multiple scattering within the medium, is 
a powerful tool, that relates arbitrary inputs to their outputs, and in principle allows the reconstruction/prediction of either.
For many years it has been understood that elastic multiple scattering can lead to Anderson 
localization of waves, and that even in the diffusive regime it creates important correlations in the scattering matrix of a random medium \cite{FKLS}.  
In electron transport, the input electron state is uncontrolled and only a few statistical properties of the scattering matrix can be measured, through, e.g., conductance and shot noise experiments, 
but in classical wave systems it is possible to prepare specific input states and measure approximately the full $S$-matrix. 
There has been a great deal of interest in
doing this in optical systems recently, since this knowledge allows the synthesis of input states, using spatial light modulators \cite{vanPutten10}, which provide dramatic control of the transmitted and reflected waves. Most strikingly one can focus light within or at the output of a strong scattering 
medium \cite{vellekoop08, popoff10b}, with potential applications for imaging biological tissue, enhancing the sensitivity of spectroscopy and, potentially, allowing the transmission of information through media which are opaque to typical input states \cite{mosk12, cizmar12}. 

For most cases under experimental study, the $S$-matrix can be naturally divided into blocks containing transmission and reflection matrices, $t$ and
$r$, with 
a certain number of input and output channels, $N$.
The possibility of strongly enhanced transmission of waves through multiple scattering media was first discovered theoretically more than twenty years ago
\cite{dorokhov84, imry86, beenakker97}, when it was shown that the eigenvalues, $T_n$, of the Hermitian matrix $t^\dagger t$ have a bimodal distribution consisting of
a large number of strongly reflected ``closed" eigenchannels, and $G$ ``open" eigenchannels with $T_n \simeq 1$ (where $G=N\ell/L$ is the dimensionless
conductance, $L$ is the sample length and  $\ell \ll L$ is the elastic mean free path). If one were able to prepare the coherent superposition of channel states corresponding to an open eigenchannel, the input state would be transmitted with near unit efficiency through
an effectively opaque medium ($\bar{T} = \ell/L \ll1$). A distinct but related effect, discovered more recently, is coherent enhancement of absorption (CEA) \cite{chong11}, the possibility of preparing an input state which is very strongly absorbed in a ``white" medium with a $\ell \ll L$ and, $\ell_a $, the inelastic
absorption length, greater than $L$.

Despite these exciting predictions of theory, experimental efforts have as yet been unable to measure these effects.
In particular, measurements of the transmission eigenvalue density have not revealed the predicted bimodal structure. The eigenvalues of
the Hermitian matrix $t^\dagger t$ are the square of the singular values, $\sqrt{T_n}$, of the complex matrix $t$, the 
distribution of which was first measured in acoustics \cite{sprik08, aubry09} and then in optics \cite{popoff10a}, with results very close to the ``quarter-circle law" \cite{marchenko67},
characteristic of {\it uncorrelated} Gaussian random matrices. The discrepancy with the theoretical prediction was attributed to the incomplete angular coverage of the input and output channels in the experiments \cite{vanPutten10}. Efforts have been made to measure a larger fraction of the transmission matrix in order to reveal the existence of open eigenchannels \cite{kim12}, but strikingly, even in experiments performed with microwaves in a multimode waveguide, where almost all channels can be addressed, the measured distribution does not reveal the second peak associated with completely open eigenchannels \cite{shi12}. 

These observations motivate the study of a new random matrix ensemble, directly relevant to the experimental measurements, which
we refer to as a filtered random matrix (FRM) ensemble. The definition of this ensemble and many of our results are quite general, and would
apply to $S$-matrices in arbitrary scattering geometries, and more generally, to any situation in which only a portion of a physical random matrix, $A$,
is measurable; however here we apply the theory to the important case of random transmission and reflection matrices.
In almost all such experiments there are some limitations on generating input channel states
(e.g. due to the limited numerical aperture associated with the incoming light) and in detecting the outgoing states; we will refer to both situations
as incomplete channel control (ICC).  Hence the experiments will typically not have access to the full, $N \times N$, matrices, $t,r$, but rather
to some finite sub-matrix of $t,r$.  We derive below the statistical properties of such measured sub-matrices, to assess the effects of ICC
on the correlations inherent in diffusive transmission and reflection.

The effect of ICC is to map the $N \times N$ matrix $A$ to $\tilde{A} = P_2AP_1$, where $P_1,P_2$ are $N \times M_1$ and $M_2 \times N$ 
matrices which eliminate $N-M_1$ columns and $N-M_2$ rows, respectively, of the original random matrix $A$. 
$\tilde{A}$ is the measured random matrix with ICC and $M_1,M_2 \leq N$ are the number of input (output) channels controlled. 
We will compute the eigenvalue density $p_{\tilde{A}^\dagger \tilde{A}}(x)$  of the Hermitian filtered random matrix, 
$\tilde{A}^\dagger \tilde{A}$, in the limit $N, M_1, M_2 \to \infty$, with arbitrary but fixed fractions of controlled channels,
$m_1=M_1/N$ and $m_2=M_2/N$, assuming that the eigenvalue density (and resolvent, see below) of the full matrix, $A^\dagger A$, is known.

\begin{figure}[t]
\centering{
\hspace{-0.3cm}
\includegraphics[angle=0,width=0.48\textwidth]{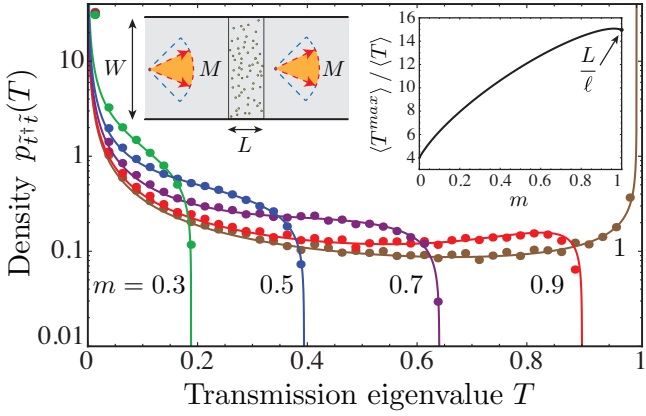}
\caption{
Transmission eigenvalue density of a disordered slab placed in a waveguide with $N=485$ channels (length $L=150/k$, width $W=900/k$), for different fractions of controlled channels $m=M/N$. Numerical results (dots) are obtained from solving the wave equation for 120 realizations of the slab, with dielectric function $\epsilon(\mathbf{r})=n_0^2+ \delta \epsilon(\mathbf{r})$; $n_0=1.5$ and $\delta \epsilon(\mathbf{r})$ is uniformly distributed between $[-1.1, +1.1]$ in the slab, and $\delta \epsilon (\mathbf{r})=0$ in the empty waveguide. The solid lines are the theoretical prediction based on Eqs.~(\ref{LinkpImG}), (\ref{SelfG}) and (\ref{ResolventT}), where $\bar{T}=\left<\sum_{n=1}^NT_n\right>/N=0.067$ is found from the simulation with complete channel control ($m=1$). Inset shows the maximal transmission enhancement possible for a given $m$, where $\left<T\right> = m \bar{T}$, and $\left<T^{\textrm{max}}\right>$ is calculated by method given in \cite{Suppl}.
\label{fig1}
}}
\end{figure}

In order to obtain the eigenvalue density $p_{\tilde{A}^\dagger \tilde{A}}(x)$, it is convenient to introduce the resolvent
\be
g_{\tilde{A}^\dagger \tilde{A}}(z)= \frac{1}{N} \left\langle \textrm{Tr} \frac{1}{z-\tilde{A}^\dagger \tilde{A}} \right\rangle,
\ee
where averaging $\left<\,\dots \right>$ denotes the ensemble or disorder average. The eigenvalue density is given by 
\be
\label{LinkpImG}
p_{\tilde{A}^\dagger \tilde{A}}(x)= - \frac{1}{\pi} \lim_{\eta \to 0^+} \mathrm{Im}
g_{\tilde{A}^\dagger \tilde{A}}(x + \textrm{i} \eta).
\ee
Although field-theoretic or diagrammatic approaches to this problem are possible, here we take advantage of the multiplicative structure of $\tilde{A}^\dagger \tilde{A}=P_1^\dagger A^\dagger P_2^\dagger P_2 A P_1$ to employ the powerful method of free probability theory \cite{voiculescu83}. This theory identifies a sufficient condition --- ``asymptotic freeness" --- under which the spectral properties of a product of matrices can be found algebraically from the spectral properties of the factors. Loosely speaking, asymptotic freeness can be thought of as the generalization of statistical independence to the case where the random ``variables" do not commute \cite{tulino04}. In our case, $A$ is assumed to be random, and the matrices $P_1$ and $P_2$ can be generated by randomly suppressing rows or columns of an $N \times N$ identity matrix.  If the ``lost" channels are not chosen randomly, the eigenvalue density is still given by the present theory with renormalized parameters, which can be determined from microscopic treatments of the radiative transfer equation \cite{Suppl}.  Here we only discuss ICC in channel (momentum) space, but we find that the FRM distributions we obtain can describe the behavior of a focused spot of light (real space filtering), incident on a waveguide or even on a slab with no walls at all \cite{goetschy13}.

Specializing general results of free probability theory to this specific ensemble, we show in \cite{Suppl} that the unknown resolvent $g_{\tilde{A}^\dagger \tilde{A}}(z)$ may be obtained from the known resolvent
$g_{A^\dagger A}(z)$ by means of the implicit equation:
\be
\label{SelfG}
N(z)\,g_{A^\dagger A}\! \left(N(z)^2/D(z)\right)=D(z),
\ee
where $N(z)$ and $D(z)$ are two auxiliary functions defined as
\begin{align}
N(z)&=z\,m_1g_{\tilde{A}^\dagger \tilde{A}}(z)+1-m_1,
\\
D(z)&=m_1g_{\tilde{A}^\dagger \tilde{A}}(z)\left[
zm_1 g_{\tilde{A}^\dagger \tilde{A}}(z)+m_2-m_1
\right].
\end{align}
We obtain the results given below by solving
this self-consistent equation in the complex z-plane numerically and taking the limit of Eq.~(\ref{LinkpImG}).  The theory also gives us explicit formulas for moments of the eigenvalue density of $\tilde{A}^\dagger \tilde{A}$ in terms of moments of $A^\dagger A$, which we use at some points in deriving the results given below \cite{Suppl}.

We now use Eq.~(\ref{SelfG}), setting $A=t$, to study the effect of  ICC on the transmission through a disordered non-absorbing slab in the diffusive regime,
in which $N\bar{T} = G > 1$, where $\bar{T}=\left<\sum_{n=1}^NT_n\right>/N\equiv \ell/L$.
The resolvent associated to the bimodal transmission eigenvalue density $p_{t^\dagger t}$ \cite{dorokhov84, nazarov94, beenakker97} is 
\be
\label{ResolventT}
g_{t^\dagger t}(z)=\frac{1}{z}-\frac{\bar{T}}{z\sqrt{1-z}}\textrm{Arctanh}\left[
\frac{\textrm{Tanh}(1/\bar{T})}{\sqrt{1-z}}\right].
\ee
The solution for the density $p_{\tilde{t}^\dagger \tilde{t}} (T)$, obtained from Eqs.~(\ref{SelfG}), (\ref{ResolventT}) and (\ref{LinkpImG}), is shown in Fig.~\ref{fig1}, where we chose $m_1=m_2\equiv m$.  For $m=1$, $p_{\tilde{t}^\dagger \tilde{t}} (z)$ has the expected bimodal shape, even for the slab geometry simulated in Fig.~\ref{fig1}, confirming that this distribution is not restricted to the quasi-one-dimensional geometry $L\gg W$ \cite{nazarov94}. The number of open eigenchannels (the channels with $T\ge1/e$) is equal to the dimensionless conductance $G=N\bar{T}$; for the most open eigenchannel, $\left<T^{\textrm{max}}\right> \to 1$ as $N \to \infty$. Introducing a small degree of ICC
$(m \lesssim 1)$ abruptly suppresses the most open eigenchannels:  the mean of the largest eigenvalue $\left<T^{\textrm{max}}\right>$ becomes strictly smaller than $1$ as $N \to \infty$, and the distribution loses its second characteristic peak. This striking property (preserved for $m_1 \neq m_2$) 
indicates that phenomena based on extremely open eigenchannels are highly sensitive to ICC and it
may explain why the bimodal shape has not been observed in real experiments, even with almost complete channel control \cite{shi12}.  
Note, however, that even if the bimodal shape is lost, $\left<T^{\textrm{max}}\right> \gg \bar{T}$ can still hold (see inset to Fig.~\ref{fig1}), for reasonable values of $m$, so strongly enhanced total transmission is not ruled out by ICC.

Our analytical prediction is in excellent agreement with the result of numerical simulations of the wave equation $[\nabla^2+k^2\epsilon(\mathbf{r})]\psi(\mathbf{r})=0$, based on numerical discretization in a two-dimensional disordered slab embedded in a multimode waveguide with $N=485$ channels and perfectly reflecting walls. The $N \times N$ transmission matrix is computed using the recursive Green's function method \cite{baranger91}, and
members of the filtered ensemble are then generated by random projection.

When $m$ is further reduced, the correlations contained in the transmission matrix are progressively lost and $p_{\tilde{t}^\dagger \tilde{t}} (T)$ evolves such that the distribution of $X=\sqrt{T/\left<T\right>}$ converges to the quarter circle law, $p_{X} (x)= \sqrt{4-x^2}/\pi$, independent of $\ell$. Thus universal, uncorrelated behavior is reached 
when $m\lesssim \bar{T}$ (or $M \lesssim G$), in agreement with measurements reported in \cite{popoff10a}.  
This loss of correlations in the limit of small degree of channel control remains when $m_1\neq m_2$, but in a more subtle form. For example, for $m_1\equiv m \lesssim \bar{T}$ and $m_2=1$, we find that $p_{X}$ approaches the Marchenko-Pastur (MP) law \cite{marchenko67}, describing rectangular random matrices with uncorrelated Gaussian matrix elements,
but for a matrix ensemble with a disorder-dependent, effective value of $m \to \tilde{m}$. Specifically
\be
\label{ProbaX}
 p_X(x)\simeq \frac{1}{\pi \tilde{m}x}\sqrt{\left(x^+-x^2\right)\left(x^2-x^-\right)},
 \ee
 where $x^\pm=\left(1\pm \sqrt{\tilde{m}}\right)^2$ and $\tilde{m}  \equiv m(2/3\bar{T}-1)$ \cite{goetschy13}; this corresponds to a MP distribution for $\tilde{N} \times M$ matrices, with $\tilde{N}=3\bar{T}N/(2-3\bar{T})$.

\begin{figure}[t]
\centering{
\hspace{-0.3cm}
\includegraphics[angle=0,width=0.48\textwidth]{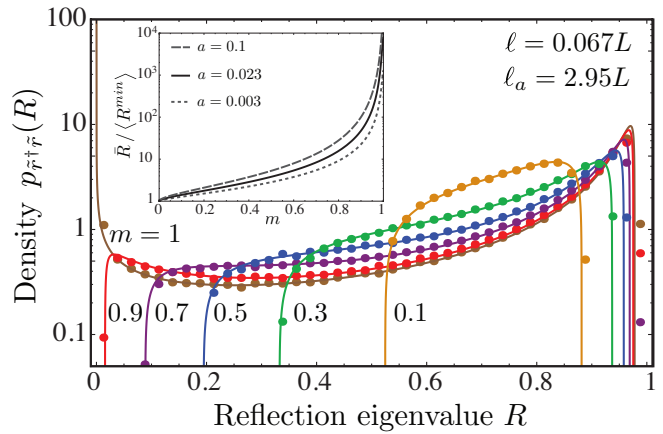}
\caption{
Reflection eigenvalue density of a disordered absorbing slab with the same geometry and dielectric function as in Fig.~\ref{fig1} except for the addition to $\epsilon (r)$ of a constant
imaginary part, $ 0.003\textrm{i}$, representing absorption.  
A fraction $m=M/N$ of the input channels are excited while all output reflection channels are collected. Numerical results are based on $100$ realizations of the 
disordered slab. Solid lines are the theoretical prediction based on Eqs.~(\ref{LinkpImG}), (\ref{SelfG}) and (\ref{ResolventR}); where $a=0.023$ is determined by the 
numerical value of $\bar{R} = 0.74$ with $m=1$. The ballistic and diffusive absorption lengths are $\ell_a=\ell/a=2.95 L$ and $\sqrt{\ell\ell_a}=0.44L$.  
\label{fig2}
}}
\end{figure}

The statistics of lossless reflection with ICC can be obtained similarly to those of transmission, with qualitatively similar results (i.e. suppression of extremal values and convergence to an effective MP distribution).
However, in the case of an {\it absorbing} disordered medium within a waveguide, one has a distinct statistical ensemble \cite{beenakker96, bruce96} from that in the lossless case. The extremal eigenvalue statistics of this ensemble were recently studied by Chong and Stone \cite{chong11} and lead to the phenomenon of coherently enhanced absorption (CEA). Here the (non-unitary) $S$-matrix and the reflection matrix coincide, and $1-R_n$ represents the absorbed fraction of an incident eigenchannel, where $\{R_n\}$ are the eigenvalues of $r^\dagger r$.  Let $\ell_a \gg \ell$ be the ballistic absorption length and consider the regime in which 
$\ell \ll \sqrt{\ell \ell_a} < L, \ell_a$, so that elastic scattering is strong, absorption is weak, but transmission is negligible. In Ref.~\cite{chong11} it was found that
when $N^2 (\ell/\ell_a) \gg 1$, 
the smallest $R_n$ (reflectivity of the most highly absorbed eigenchannel)  was orders of magnitude smaller than the mean reflectivity, $\bar{R}$:
\be 
\frac{\left<R^{\textrm{min}}\right>}{\bar{R}} \simeq \frac{1}{2N^2a}, \;\;\;\; a \equiv \frac{\ell}{\ell_a} \ll 1.
\label{RminN}
\ee
As $N \to \infty, \left<R^{\textrm{min}}\right> \to 0$ while $\bar{R}$ remains $\sim 1$, which would not be true, e.g. for the MP law, and is the essence
of CEA; in addition the density $p_{r^\dagger r}(R)$ diverges at $R=0$ (see Fig.~\ref{fig2}).
The $1/N^2$ scaling of $\left<R^{\textrm{min}}\right>$ holds even when the absorption is non-uniform, e.g. for a buried absorber behind an ``opaque", lossless layer. 
The effect of ICC in this case is again
found by solving Eq.~(\ref{SelfG}), now with $A=r$. We will specialize to the case where a fraction $m=M/N$ of the input channels can be excited, while the field in all output channels is collected, $m_1=m$, $m_2=1$. We find the eigenvalue density $p_{\tilde{r}^\dagger \tilde{r}}$, from the known density
$p_{r^\dagger r}$\cite{beenakker96,bruce96} and the associated resolvent,
\be
\label{ResolventR}
g_{r^\dagger r}(z)=\frac{z-1+2a-2a\sqrt{1+1/a-1/az}}{(1-z)^2}.
\ee
Our results (Fig.~\ref{fig2}) again show excellent agreement with the simulation of the relevant wave equation for any $m$. Of particular interest
is the behavior and support of the density near $R=0$.  If we take the limit $N \to \infty$ and then consider $m = 1- \delta$, we find that the density has
no support at $R=0$, but instead the support has a sharp cut-off at $R \neq 0$, which we can identify with $\left<R^{\textrm{min}}\right>$.  A general equation for $\left<R^{\textrm{min}}\right>$ is 
derived in \cite{Suppl}, and its solution, normalized by $\left< R\right> = \bar{R}$, is plotted in the inset to Fig.~\ref{fig2}. The strong sensitivity of $\left<R^{\textrm{min}}\right >$ to $N$ is
completely lost for all $m$ such that $\delta > 2\sqrt{2}/N$, and instead we find to leading order,
\be
\left<R^{\textrm{min}}\right>\simeq\frac{\delta^2}{16a}
\label{Rminm}
\ee
for $\delta \ll \sqrt{a}$,
and $\left<R^{\textrm{min}}\right>\simeq \delta -3(1-\delta)^{1/3}\delta^{2/3}a^{1/3}$ for $\delta \gg a^{1/3}$.
The experimentally observable decrease in reflectivity relative to $\bar{R}$ will in most cases not be determined by $N$,
but instead will be controlled by $\delta$ and typically will be much less than predicted by Eq.~(\ref{RminN}). Note, however, that it can still be
substantial, e.g. a factor of $\sim 5$ decrease in reflectivity, when only half of the channels are controlled for realistic experimental parameters.

The joint probability distribution corresponding to this system, for $m=1$, coincides with the Gibbs distribution of a Coulomb gas of charges with coordinates $R_n$, in the presence of an external potential $u_1(R_n)$ that depends on $a$ \cite{beenakker96,bruce96, Suppl}. 
The dramatic change in the support of the distribution $p(R)$ at $R=0$ from infinite to zero when $m<1$ is related to a zero-temperature phase transition in this Coulomb gas as $m$ is decreased 
from unity \cite{Suppl}. A similar transition happens for $p(T)$ near unity for the non-absorbing system. In addition to the change
in support of the eigenvalue density, we also find \cite{Suppl} that for both the absorbing reflection and lossless transmission cases
ICC not only modifies $u_1$, but also changes the short-range correlations of the eigenvalues, inducing a crossover from linear ($\beta=1$) eigenvalue repulsion to quadratic
($\beta=2$) eigenvalue repulsion, similar to that normally associated with time-reversal (TR) symmetry breaking.  This is due to the fact that
randomly suppressing rows or columns of the complex matrices $t,r$ leads to an ensemble of $S$-matrices which violate the usual TR symmetry
constraint, $SS^*=1$. This suggests that in many experiments with nominal TR symmetry the $T$ and $R$ spectra will nonetheless show quadratic
eigenvalue repulsion, and there is some evidence to this effect \cite{Zhou}.  Finally, for $m\lesssim a$, the density of normalized absorption of each eigenchannel,
$X=\sqrt{(1-R)/(1-\bar{R})}$ is of the uncorrelated form (\ref{ProbaX}) with $\tilde{m}=m(1+2a)/4\sqrt{a(1+a)}-m/2$ \cite{goetschy13}.

The previous analyses suggests that the degree of correlations contained in the matrices $\tilde{t},\tilde{r}$ is controlled by
the parameters $m/\bar{T},m/a$, so that when the fractional control is less than the ``loss rates" $\bar{T}$ and $a$, correlations are lost.
To make this statement precise in the sense of information theory, we have studied the information capacity, $C$, of a disordered multimode waveguide, focusing on the case of transmission without absorption.  The information capacity $C$ is the maximal rate, expressed in bits per second per Hertz (bps/Hz), at which the sender can transfer information with a vanishingly low probability of error \cite{cover91}. Our microscopic theory for $p_{\tilde{t}^\dagger\tilde{t}}(x)$ allows us to compute $C$ for arbitrary choice of $m_1$ and $m_2$  \cite{goetschy13}. First, for $m_1=m_2\equiv m =1$, we find $C=G\,\textrm{ln}^2
\left(
\sqrt{1+\textrm{\footnotesize{SNR}}/\bar{T}} 
+  
\sqrt{\textrm{\footnotesize{SNR}}/\bar{T}}
\right)/\textrm{ln}2$, where $\textrm{\footnotesize{SNR}}$ is the signal to noise ratio measured at the output. This shows that (up to logarithmic corrections) the number of open eigenchannels $G$ can be interpreted as the bitrate of the disordered sample with complete channel control. Second, in the regime of strong ICC, $m\lesssim \bar{T}$, the capacity per channel increases, becoming independent of $\bar{T}$, and is given by
$C/M=2\textrm{log}_2(1+\sqrt{1+4\,\textrm{\footnotesize{SNR}}})-2-1/\textrm{ln}2+\left(\sqrt{1+4\,\textrm{\footnotesize{SNR}}}-1 \right)/2\textrm{ln}2\,\textrm{\footnotesize{SNR}}$. 
This is the standard form used to model free space communication where many channels are uncontrolled \cite{tulino04}. In the intermediate regime, $\bar{T}<m<1, \bar{T} < 0.1$, $C/M$ depends only on the ratio $M/G$, confirming that this ratio is a measure of the degree of correlations.  As long as $M>G$ the disorder-induced correlations are revealed and the capacity per channel drops from its maximum, uncorrelated value.  

In summary, the extremal eigenvalue properties necessary for transmission through opaque media or enhanced absorption are suppressed substantially as the degree of channel control is reduced, however strong enhancements should still be possible for achievable values of the channel control parameters, $m_1,m_2$.  In most cases, experiments will need to measure $m_1$ and $m_2$ in order to estimate the maximum enhancements possible for a given system and set-up. Note that if polarization is not preserved in the scattering process and only one polarization is controlled/detected, then the parameters $m_1$, $m_2$ are immediately reduced by a factor of $1/2$.

This research was partially supported by NSF grant ECCS 1068642. We acknowledge helpful conversations with Sebastien Popoff, Hui Cao and
Zhou Shi.


\end{document}


\title {\LARGE Supplementary Material}
\maketitle

\section{Singular value density of a filtered random matrix}

Let us consider a $N\times N$ random matrix $A$ such that the eigenvalue density $p_{A^\dagger A}(x)$ of the Hermitian matrix $A^\dagger A$ is known. Since the singular values of $A$ are the square root of the eigenvalues $\Lambda$ of $A^\dagger A$, the singular value density of $A$ is $p^{SV}_{A}(x)=2xp_{A^\dagger A}(x^2)$. Our goal is to show how the singular spectrum of $A$ is modified when we suppress a certain fraction of rows and columns in the latter.

The modified spectrum is the one of a $M_2\times M_1$ submatrix of $A$ ($M_1,M_2<N$) that can be represented as $\tilde{A}=P_2AP_1$, where $P_1$ and $P_2$ are two rectangular matrices that select $M_1$ columns and $M_2$ rows of $A$, respectively.    
In order to find the eigenvalue density of $\tilde{A}^\dagger\tilde{A}=P_1^\dagger A^\dagger P_2^\dagger P_2 AP_1$, we introduce the resolvent
\be
\label{Defresolvent}
g_{A^\dagger A}(z) = \frac{1}{N} \left\langle \textrm{Tr} \frac{1}{z-A^\dagger A} \right\rangle.
\ee
The eigenvalue density of $A^\dagger A$,
\be
p_{A^\dagger A}(x)=\frac{1}{N}\left\langle \sum_{n=1}^N\delta(x-\Lambda_n)
 \right\rangle,
\ee
is related to the resolvent through
\be
\label{LinkpImG}
p_{A^\dagger A}(x)= - \frac{1}{\pi} \lim_{\eta \to 0^+} \mathrm{Im}
g_{A^\dagger A}(x + \textrm{i} \eta),
\ee
or, inversely,
\be
\label{LinkGp}
g_{A^\dagger A}(z)=\int_{-\infty}^\infty\textrm{d}x\frac{p_{A^\dagger A}(x)}{z-x}.
\ee
The combination of the definition (\ref{Defresolvent}) with the relation $\textrm{Tr}(z-BC)^{-1}=\textrm{Tr}(z-CB)^{-1}+(M-N)/z$ (valid for arbitrary matrices $B$ and $C$ of size $M\times N$ and $N \times M$, respectively) yields
\be
\label{GAdaggerA}
g_{\tilde{A}^\dagger\tilde{A}}(z)=\frac{g_{A^\dagger \mathcal{P}_2A\mathcal{P}_1}(z)}{m_1}-\frac{1-m_1}{m_1z},
\ee
where $m_1=M_1/N$, and $\mathcal{P}_1=P_1P_1^\dagger$ and $\mathcal{P}_2=P_2^\dagger P_2$ are two projectors of size $N\times N$. The problem is then reduced to finding the resolvent of the matrix $A^\dagger \mathcal{P}_2A\mathcal{P}_1$.

As explained in the main text (MT), our idea is to use free probability theory \cite{voiculescu83, tulino04}. The eigenvalue distribution of the product of two asymptotically free matrices $X_1$ and $X_2$ can be found using the so-called  $\mathcal{S}$-transform \cite{tulino04}.  
If we define $\chi(z)$ as a solution of 
\be
\label{DefchiRMT}
\frac{1}{\chi(z)}g\left(\frac{1}{\chi(z)}\right)-1=z,
\ee
then the $\mc{S}$-transform is
\be
\label{DefSRMT}
\mc{S}(z)=\frac{1+z}{z}\chi(z).
\ee
The $\mc{S}$-transform of the product $X_1X_2$ satisfies \cite{tulino04}:
\be
\label{Sproduct}
\mc{S}_{X_1X_2}(z)=\mc{S}_{X_1}(z)\mc{S}_{X_2}(z).
\ee
Since the projectors $\mathcal{P}_1$ and $\mathcal{P}_2$ are not random, we will assume that they are asymptotically free with respect to the random matrix $A^\dagger A$, in the limit $N,M_1,M_2\to \infty$ with $m_1=M_1/N$ and $m_2=M_2/N$ fixed. Similarly, we assume that $A^\dagger \mathcal{P}_2A$ and $\mathcal{P}_1$ are asymptotically free. With the property (\ref{Sproduct}), we obtain 
 \be
 \label{SAPA}
 \mc{S}_{A^\dagger \mathcal{P}_2A\mathcal{P}_1}(z)= \mc{S}_{ \mathcal{P}_2}(z)\mc{S}_{A^\dagger A}(z)\mc{S}_{ \mathcal{P}_1}(z),
 \ee
where we used $\mc{S}_{A^\dagger \mathcal{P}_2A}(z)=\mc{S}_{AA^\dagger \mathcal{P}_2}(z)$ and $\mc{S}_{AA^\dagger}(z)=\mc{S}_{A^\dagger A}(z)$. In addition, the projector $\mathcal{P}_1$ ($\mathcal{P}_2$) has $M_1$ ($M_2$) eigenvalues equal to one and $N-M_1$ ($N-M_2$) eigenvalues equal to zero. Therefore, $g_{\mathcal{P}_{1,2}} (z)=m_{1,2}/(z-1)+(1-m_{1,2})/z$. According to the definitions (\ref{DefchiRMT}) and (\ref{DefSRMT}), the $\mathcal{S}$-transform of the projectors are:
\be
\label{SProjector}
\mc{S}_{ \mathcal{P}_{1,2}}(z)=\frac{z+1}{z+m_{1,2}}.
\ee
Finally, combining Eqs.~(\ref{SAPA}) and (\ref{SProjector}), and after some algebra using Eqs.~(\ref{GAdaggerA}), (\ref{DefchiRMT}) and (\ref{DefSRMT}), we obtain Eq.~(3) of the MT, which is an implicit equation obeyed by the unknown resolvent $g_{\tilde{A}^\dagger\tilde{A}}(z)$. If the singular value density of $A$ is known, the singular value density of $\tilde{A}$ is obtained by applying the following recipe. (1) Compute the resolvent of $A^\dagger A$ with Eq.~(\ref{LinkGp}). (2) Find (numerically) the solution $g_{\tilde{A}^\dagger\tilde{A}}(z)$ of Eq.~(3) of the MT. (3) Deduce the singular value density $p^{SV}_{\tilde{A}}(x)=2xp_{\tilde{A}^\dagger\tilde{A}}(x^2)$ from Eq.~(\ref{LinkpImG}).

The different properties of the solution of the equation obeyed by  $g_{\tilde{A}^\dagger\tilde{A}}(z)$ will be detailed elsewhere \cite{goetschy13}.  Here, we simply mention some properties implicitly used throughout the Letter:
\begin{itemize}
\item 
Since $\tilde{A}$ is of size $M_2\times M_1$, the $M_1\times M_1$ matrix $\tilde{A}^\dagger\tilde{A}$ is of rank $\textrm{min}(M_1,M_2)$, so that  $\tilde{A}^\dagger\tilde{A}$ has $\textrm{max}(M_1-M_2,0)$ eigenvalues equal to zero.  
\item
The eigenvalue density of $\tilde{A}\tilde{A}^\dagger$ is simply related to the one of  $\tilde{A}^\dagger\tilde{A}$ by
\be
p_{\tilde{A}\tilde{A}^\dagger}(x)=\frac{m_1}{m_2}p_{\tilde{A}^\dagger\tilde{A}}(x)+\left(1-\frac{m_1}{m_2}\right)\delta(x).
\ee
\item
The first two cumulants of the  distribution $p_{\tilde{A}^\dagger\tilde{A}}$ are
\begin{align}
\label{Moment1}
\left<\Lambda_{\tilde{A}^\dagger\tilde{A}}\right>&=m_2\left<\Lambda_{A^\dagger A}\right>,
\\
\textrm{Var}\Lambda_{\tilde{A}^\dagger\tilde{A}}&=m_1m_2\left[
m_2\textrm{Var}\Lambda_{A^\dagger A}
+(1-m_2)\left\langle\Lambda_{A^\dagger A}\right\rangle^2
\right].
\end{align}
\item
It can also be shown that the edges $x^*$ of the distribution are given by
\be
\label{SolEdge1}
 x^*=\xi^*\left[1+\frac{m_1-1}{\xi^*g_{A^\dagger A}(\xi^*)}\right]\left[1+\frac{m_2-1}{\xi^*g_{A^\dagger A}(\xi^*)}\right] ,
 \ee
where $\xi^*$ is the solution of
\begin{align}
\label{SolEdge2}
&\left.\frac{\textrm{d}g_{A^\dagger A}(\xi)}{\textrm{d}\xi}\right\vert_{\xi^*}=\frac{g_{A^\dagger A}(\xi^*)}{2\xi^*} \, \times
\\
\nonumber
&\frac{-(1-m_1)(1-m_2)+\xi^{*2}g_{A^\dagger A}(\xi^*)^2}{(1-m_1)(1-m_2)-(1-m_1/2-m_2/2)\xi^*g_{A^\dagger A}(\xi^*)}.
\end{align}
\end{itemize}

Equations (\ref{SolEdge1}) and (\ref{SolEdge2}) have been used to obtain, in the limit $N\to \infty$, $\left<T^{max}\right>$ without absorption (see inset to Fig.~1 of the MT) and  $\left<R^{min}\right>$ with absorption (see inset to Fig.~2 of the MT). In order to illustrate the accuracy of those predictions in the context of coherent enhancement of absorption, $\left<R^{min}\right>$, evaluated from Eqs.~(\ref{SolEdge1}), (\ref{SolEdge2}) and Eq.~(9) of the MT, is compared in Fig.~1 with the result of numerical simulations of the wave equation, for different values of absorption.

\begin{figure}[t]
\centering{
\hspace{-0.2cm}
\includegraphics[angle=0,width=0.48\textwidth]{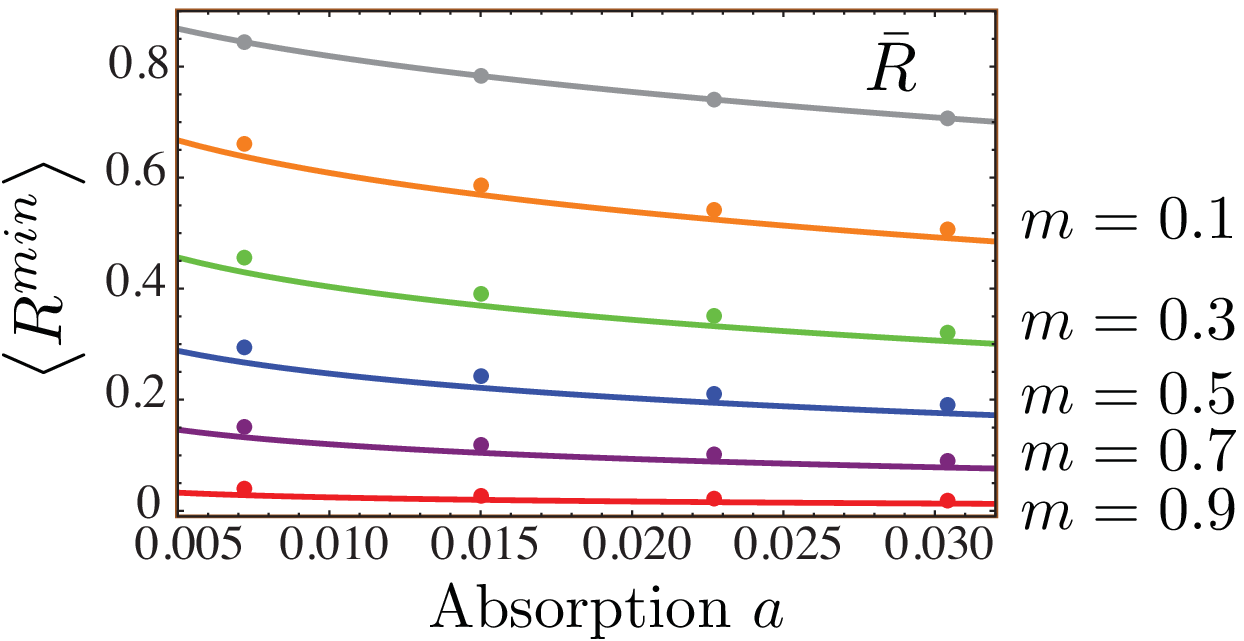}
\caption{
Variation of the smallest reflection eigenvalue $\left<R^{min}\right>$ with $m_1\equiv m=M/N$ and $a=\ell/\ell_a$. Each data point is an average over $100$  realizations, with the same parameters as in Fig.~2 of the MT; $a$ is determined by the numerical value of $\bar{R}=\left<\sum_{n=1}^NR_n\right>/N=1+2a-2\sqrt{a(1+a)}$, and solid lines follow from the equations for the edge of the distribution $p_{\tilde{r}^\dagger\tilde{r}}$ (see text).
\label{fig1}
}}
\end{figure}

\section{Filtering the transmission and reflection matrices with a non-random subset of channels}

In a diffusive sample, waves lose the memory of their initial direction after a distance of the order of the transport mean free path $\ell_{tr}$ \footnote{$\ell_{tr}$ differs by a numerical factor from the mean free path usually defined in random matrix theory and used in the MT, $\ell=\bar{T}L$ \cite{beenakker97}.}, meaning that, inside the sample, all momentum channels diffuse in the same way. This observation justifies the fact that the projectors $ \mathcal{P}_1$ and $ \mathcal{P}_2$ should be asymptotically free with respect to $\tilde{t}^\dagger \tilde{t}$ and  $\tilde{r}^\dagger \tilde{r}$. However, since the different incoming channels are not converted into diffusons at the same depth inside the sample, boundary conditions make the channels slightly non-equivalent. To illustrate this point, let us consider a waveguide with $N$ propagating channels. A typical experimental constraint might be that all the channels in
a given angular range (numerical aperture) are controlled on incidence and collected on output, so the ``lost channels" are not chosen randomly.
Using the radiative transfer equation for the specific intensity of light, it can be shown that an incoming channel oriented with an angle $\theta_a$ with respect to the longitudinal axis of the waveguide is transmitted in the outgoing direction $\theta_b$ with an amplitude $\left<T_{ab}\right>=\left<\vert t_{ab}\vert^2\right>$ of the form \cite{nieuwenhuizen93, akkermans06}
\be
\left<T_{ab}\right>=\chi_a\chi_b\frac{\bar{T}}{N},
\ee
where $\chi_a=\left<T_{a}\right>/\bar{T}=\sum_{b=1}^N\left<T_{ab}\right>/\bar{T}$. 
 In 2D, $\chi_a=2[\cos(\theta_a)+z_0/\ell_{tr}]/\pi$, and in 3D, $\chi_a=3[\cos(\theta_a)+z_0/\ell_{tr}]/4$. Here, $z_0$ is the extrapolation length, equal to $\pi \ell_{tr}/4$ in 2D and $2 \ell_{tr}/3$ in 3D, in the absence of internal reflections at the boundaries of the medium \cite{akkermans06}. It then follows that the mean value of the distribution $p_{\tilde{t}^\dagger \tilde{t}}$ is given by
\be
\label{MeanTMicro}
\left<T\right>=\frac{1}{M_1}\sum_{a=1}^{M_1}\sum_{b=1}^{M_2}\left<T_{ab}\right>=m_2s_{\{b\}}(m_2)s_{\{a\}}(m_1)\bar{T},
\ee
where $s_{\{a\}}(m)=\sum_{a=1}^M\chi_a/M$ satisfies $s_{\{a\}}(1)=1$. If the subset ${\{a\}}$ is chosen randomly among the $N$ channels, we get  $s_{\{a\}}(m)=1$ for arbitrary $m$ and $\left<T\right>=m_2\bar{T}$, in agreement with Eq.~(\ref{Moment1}). The latter property is not exactly satisfied for a non-random subset ${\{a\}}$. If we select in 2D the $M$ channels with angles $\vert \theta \vert<\textrm{Arcsin}(m)$, we would get $s(m)=m/2+\textrm{Arcsin}(m)/\pi+m\sqrt{1-m^2}/\pi>1$ and $\left<T\right>>m_2\bar{T}$.

The effect of the surface layers where free propagation is converted into diffusion can be taken into account by a renormalization of the ICC parameters. Equation (\ref{MeanTMicro}) suggests to renormalize the parameters $m_2$ and $\bar{T}$ as:
\be
\label{Renormalize}
m_2 \to m_2s_{\{b\}}(m_2), \;\;\; \bar{T} \to s_{\{a\}}(m_1)\bar{T}.
\ee
To test the validity of our theory, we have reevaluated the transmission and reflection eigenvalue distributions studied in Figs.~1 and  2 of the MT with projectors corresponding to a limited numerical aperture $\vert \theta \vert<\textrm{Arcsin}(m)$, and compared them with ICC predictions. When we use the bare values of $m_2$ and $\bar{T}$, a shown in Fig.~2, one still finds reasonable agreement, but with small deviations from numerical data (solid lines).
These discrepancies are nicely corrected by the renormalization (\ref{Renormalize}) (dashed lines), confirming that their microscopic origin is the 
surface layer effect studied above.

\begin{figure}[t]
\centering{
\hspace{-0.2cm}
\includegraphics[angle=0,width=0.48\textwidth]{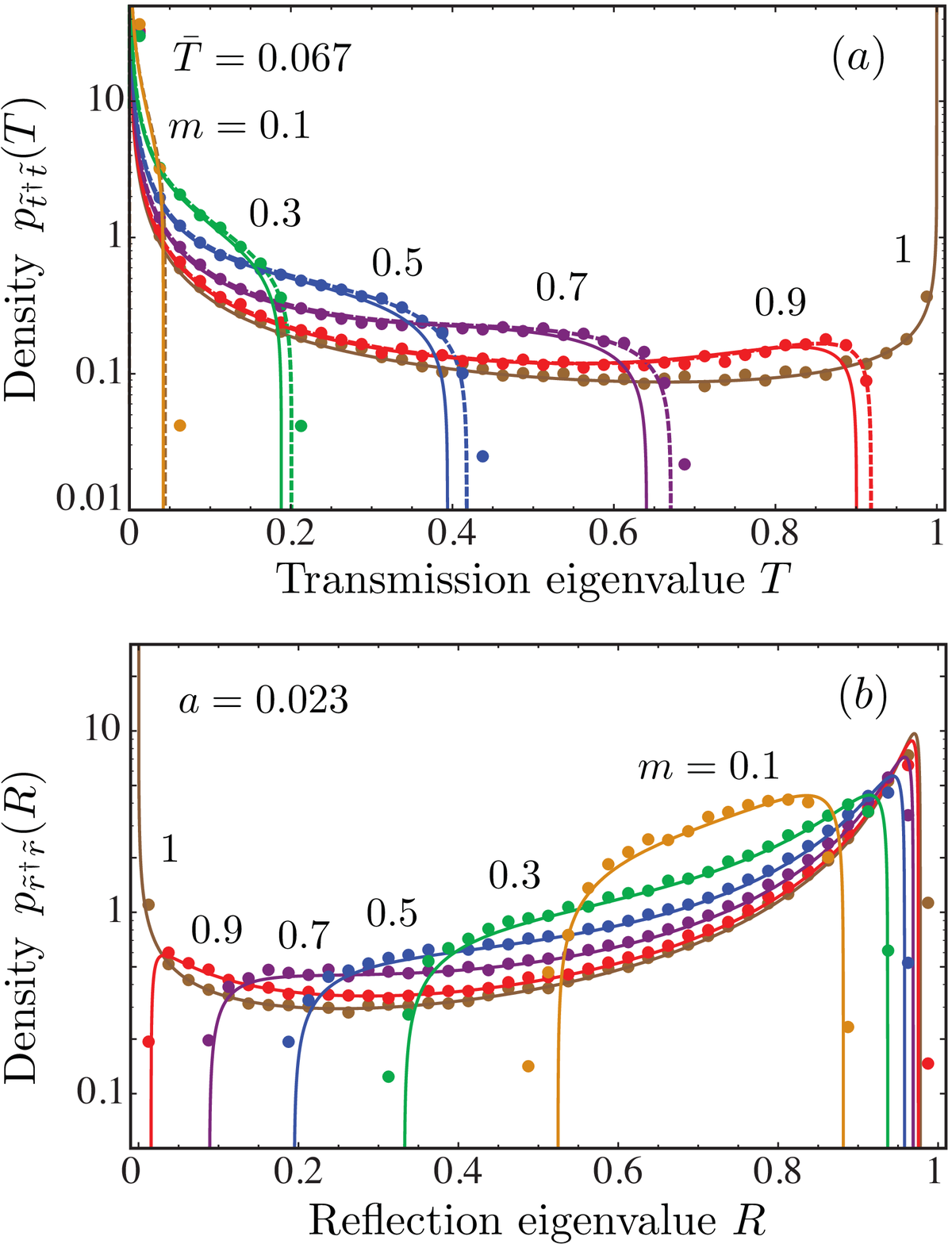}
\caption{
Effect of a limited numerical aperture on the measured transmission and reflection eigenvalue densities. Numerical results (dots) in (a) and (b) are obtained for the same parameters as in Figs.~1 and 2 of the MT, respectively. But here the projectors $P_1$ and $P_2$ are non-random and mimic a numerical aperture $\vert \theta \vert<\textrm{Arcsin}(m)$. (a) $m_1=m_2=m$. Solid lines are identical to Fig.~1 of the MT, and dashed lined correspond to ICC theory with renormalized parameters (\ref{Renormalize}); $s(m)=\sum_{a=1}^M\left<T_a\right>/M\bar{T}$ is determined from the numerical $t$-matrix with $m=1$. (b) $m_1=m$, $m_2=1$. No significant deviation is observed with respect to the results presented in Fig.~2 of the MT.
\label{fig2}
}}
\end{figure}

\section{Phase transition in the Dyson gas}
 As explained in the MT, we observe a dramatic change in the support of the distribution $p_{\tilde{t}^\dagger \tilde{t}}(T)$ at $T=1$ from infinite to zero when $m<1$. A similar transition happens for $p_{\tilde{r}^\dagger \tilde{r}}(R)$ at $R=0$, with or without absorption. To get more insight in this transition, let us first consider the case of an absorbing quasi-1D disordered waveguide. 
For $m_1=m_2=1$ and $L>\sqrt{\ell\ell_a}$, the joint probability density (JPD) of the reflection eigenvalues $R_n\in[0,1]$ is of the form  \cite{beenakker96, bruce96}
\be
\label{Gibbs}
P(\{R_n\})\varpropto e^{-\beta\left[\sum_n u_1(R_n)+\sum_{n<m}u_2(R_n,R_m)\right]},
\ee
where
\begin{align}
\label{Potential1}
u_1(R_n)&=(N+1)\left[2a\frac{R_n}{1-R_n}+\textrm{ln}\left(1-R_n\right)\right],
\\
\label{Potential2}
u_2(R_n,R_m)&=-\textrm{ln}\vert R_n - R_m \vert,
\end{align}
 and $\beta=1$ (unbroken time-reversal symmetry). The JPD coincides with the Gibbs distribution of a two-dimensional Coulomb gas with coordinates $R_n$ confined to the segment $[0,1]$ and experiencing the one-body potential $u_1$. The Coulomb repulsion tends to spread the eigenvalues on the real line but because of the ``hard wall'' at $R=0$,  they accumulate in the vicinity of the latter, giving rise to the peak of open eigenchannels observed near $R=0$ in Fig.~2.
 On the other hand, for  $m_1\equiv m\lesssim a$ and $m_2=1$, the density of $X=\sqrt{(1-R)/(1-\bar{R})}$ is of the Marchenko-Pastur type [see Eq.~(7) of the MT] with $\tilde{m}=m(1+2a)/4\sqrt{a(1+a)}-m/2$ \cite{goetschy13}. The JPD of the $M$ eigenvalues corresponding  to this law is still a Gibbs distribution, with \cite{james64, marchenko67}
\be
\label{Potential1MP}
u_1(R_n)=\frac{M}{2\tilde{m}}\left[\frac{1-R_n}{1-\bar{R}}+(\tilde{m}-1)\textrm{ln}\left(1-R_n\right)\right],
\ee
$u_2$ unchanged, and $\beta=2$ (broken time-reversal symmetry). In the intermediate regime $a\lesssim m \lesssim1$, it is reasonable to think that the potential $u_1$ evolves between the two limits (\ref{Potential1}) and (\ref{Potential1MP}), so that the modification in $\left<R^{min}\right>$ can be interpreted as a zero-temperature phase transition in the associate Coulomb gas, occurring  at $m=1$ and due to a modification of the potential induced by imperfect channel control.

A similar analysis can be performed for $\left<T^{max}\right>$ in a quasi-1D disordered waveguide. For $m_1=m_2=1$, the JPD of the transmission eigenvalues has still the form of the Gibbs distribution (\ref{Gibbs}) with 
\begin{align}
\label{Potential3}
u_1(T_n)&=\frac{N+1}{2}\left[ \ln T_n +\bar{T} x_n^2 \right],
\\
\label{Potential4}
u_2(T_n,T_m)&=-\frac{1}{2}\left[\ln \vert T_n - T_m \vert +\ln \vert x_n^2 - x_m^2 \vert\right],
\end{align}
$x_n=\textrm{tanh}^{-1}\sqrt{1-T_n}$, and $\beta=1$. The eigenvalue repulsion (\ref{Potential4}) coincides with the Coulomb interaction (\ref{Potential2}) only for the most open eigenchannels $T_n\to 1$, while for the closed  eigenchannels $T_n\to 0$, the interaction is twice as small \cite{beenakker97}. When the fraction $m_1\equiv m$ of input channels is reduced, some eigenvalues are lost and the remaining eigenvalues redistribute on the real line. In the limit $m\lesssim \bar{T}$, we find that the distribution of $X=\sqrt{T/\left<T\right>}$ is given by Eq.~(7) of the MT with $\tilde{m}=m(2/3\bar{T}-1)$, suggesting that $\tilde{t}^\dagger \tilde{t}$ has the same spectral properties as a complex Wishart matrix. The JPD of the latter is of the form (\ref{Gibbs}) with
\begin{align}
\label{Potential5}
u_1(T_n)&=\frac{M}{2\tilde{m}}\left[\frac{T_n}{\bar{T}}+(\tilde{m}-1)\textrm{ln}T_n\right],
\\
\label{Potential6}
u_2(T_n,T_m)&=-\ln \vert T_n - T_m \vert,
\end{align}
and $\beta=2$. The comparison of Eqs.~(\ref{Potential3}) and (\ref{Potential4}) with Eqs.~(\ref{Potential5}) and (\ref{Potential6}) suggests that ICC induces a modification of both $u_1$ and $u_2$, in such a way that a phase transition --- responsible for the abrupt change in $\left<T^{max}\right>$ --- occurs at $m=1$ and the usual Coulomb interaction is restored for $m\lesssim \bar{T}$. As noted in the MT, ICC is also characterized by a rapid transition $\beta=1\to \beta=2$, that has a clear signature in the  unfolded level spacing distribution: we observed numerically that the latter, given by the Wigner surmise typical of GOE for $m_1=m_2=1$,  obeys the Wigner surmise for GUE in the case of ICC \cite{goetschy13}.